\documentclass[conference]{IEEEtran}
\IEEEoverridecommandlockouts
\usepackage{stfloats}
\usepackage{cite}
\usepackage{amsmath,amssymb,amsfonts}
\usepackage{algorithmic}
\usepackage{algorithm}
\usepackage{graphicx}
\usepackage{textcomp}
\usepackage{array}
\usepackage{xcolor}
\def\BibTeX{{\rm B\kern-.05em{\sc i\kern-.025em b}\kern-.08em
    T\kern-.1667em\lower.7ex\hbox{E}\kern-.125emX}}
\usepackage{multirow}
\usepackage{makecell}
\usepackage{amsthm,amsmath,amssymb}
\usepackage{subfigure}
\usepackage{mathrsfs}    
\usepackage{subfig}
\usepackage{pifont}
\usepackage{gensymb}

\allowdisplaybreaks[4]

\begin{document}
\newtheorem{theorem}{Theorem}

\title{\textsc{SatFlow}: Scalable Network Planning for LEO Mega-Constellations \\

\thanks{This work was supported by the National Natural Science Foundation of
China (Grant No. 62302292) and the Fundamental Research Funds for the
Central Universities. The work of Bo Li was supported in part by a RGC RIF grant under the contract R6021-20, RGC TRS grant under the contract T43-513/23N-2, RGC CRF grants under the contracts C7004-22G, C1029-22G and C6015-23G, and RGC GRF grants under the contracts 16200221, 16207922 and 16207423.  
Yifei Zhu is the 
corresponding author.}
}

\author{\IEEEauthorblockN{Sheng Cen\IEEEauthorrefmark{1},
Qiying Pan\IEEEauthorrefmark{1},
Yifei Zhu\IEEEauthorrefmark{1}\IEEEauthorrefmark{3}, and Bo Li\IEEEauthorrefmark{4}}
\IEEEauthorblockA{\IEEEauthorrefmark{1}UM-SJTU Joint Institute, Shanghai Jiao Tong University}
\IEEEauthorblockA{\IEEEauthorrefmark{3}Cooperative Medianet Innovation Center (CMIC), Shanghai Jiao Tong University}
\IEEEauthorblockA{\IEEEauthorrefmark{4}Department of Computer Science and Engineering, Hong Kong University of Science and Technology}
\IEEEauthorblockA{
Email: cens98@sjtu.edu.cn, sim10\_arity@sjtu.edu.cn, yifei.zhu@sjtu.edu.cn, bli@ust.hk}
}

\maketitle

\date{}

\begin{abstract}
Low-earth-orbit (LEO) satellite communication networks have evolved into mega-constellations with hundreds to thousands of satellites inter-connecting with inter-satellite links (ISLs). 
Network planning, which plans for network resources and architecture to improve the network performance and save operational costs, is crucial for satellite network management. However, due to the large scale of mega-constellations, high dynamics of satellites, and complex distribution of real-world traffic, it is extremely challenging to conduct scalable network planning on mega-constellations with high performance. In this paper, we propose \textsc{SatFlow}, a distributed and hierarchical network planning framework to plan for the network topology, traffic allocation, and fine-grained ISL terminal power allocation for mega-constellations. To tackle the hardness of the original problem, we decompose the grand problem into two hierarchical sub-problems, tackled by two-tier modules. 
A multi-agent reinforcement learning approach is proposed for the upper-level module so that the overall laser energy consumption and ISL operational costs can be minimized; A distributed alternating step algorithm is proposed for the lower-level module so that the laser energy consumption could be minimized with low time complexity for a given topology. 
Extensive simulations on various mega-constellations validate \textsc{SatFlow}'s scalability on the constellation size, reducing the flow violation ratio by up to $21.0\%$ and reducing the total costs by up to $89.4\%$, compared with various state-of-the-art benchmarks.

\end{abstract}


\section{Introduction}

Emerging low earth orbit (LEO) satellite communication networks, like Starlink \cite{Starlink}, Kuiper \cite{kuiper}, and OneWeb \cite{oneweb}, serve as a great complement to traditional terrestrial communication. 
Recent satellite networks have further become  ``mega-constellations'' consisting of a large number of satellites for broader coverage. Starlink alone has more than 5,000 satellites in operation as of 2023 \cite{Starlink}. Furthermore, emerging LEO mega-constellations are equipped with satellite laser terminals for establishing inter-satellite links (ISLs), enabling direct high-throughput, low-latency inter-satellite transmission in space. Thanks to these advancements, marine and aerial communication where fiber deployment is unavailable or
expensive, as well as global real-time multimedia
communication (e.g. global online video conferencing \cite{DBLP:conf/infocom/LaiLWLXW22} and live video streaming \cite{DBLP:conf/mmsys/ZhaoP24}), could all greatly benefit from LEO mega-constellations.

Network planning for LEO mega-constellations aims at planning and provisioning resources to improve network performance and save operational costs. Although ISL-equipped mega-constellations have facilitated efficient global transmission, conducting network planning for them is non-trivial and faces the following challenges. First, \textit{constellation scalability and dynamics}.
Mega-constellations consist of hundreds to thousands of satellites with high velocities and highly-dynamic network properties. Planning for them could involve a large number of time-varying decision variables. 
Second, \textit{biased traffic distribution}. Real-world access traffic in LEO mega-constellations exhibits a biased distribution, with geographical and temporal patterns \cite{geo,temp}. Therefore, the network planning process should take into account the traffic distribution along with its temporal shift for better optimization. 
Third,  \textit{high laser energy consumption}. ISL-based communication requires turned-on lasers to support sustainable transmission, which consumes substantial energy for large-scale mega-constellations. In contrast, the energy harvested by solar panels in satellite networks is precious and constrained to the limited number and size of the solar panels, as well as the intermittent harvesting behavior due to their fast movement relative to the sun. Consequently, the issue of high energy consumption has to be carefully tackled for satellites' sustainable operation. 

Unfortunately, existing works fail to deal with these three major challenges simultaneously. 
Even though some works on satellite network planning \cite{DBLP:conf/sigcomm/ZhuGATZJ21,DBLP:conf/sigcomm/AhujaGDBGZLXZ21,DBLP:conf/noms/PremsankarGFV18} consider about real-world biased traffic patterns, their plannings are based on small and static networks, and have a poor performance when handling large-scale mega-constellations in a reasonable time. 
On the other hand, works dedicated to network planning on large-scale and dynamic mega-constellations \cite{DBLP:conf/icc/PiRWZZL22,DBLP:conf/networking/LinLLLF22,DBLP:journals/twc/MayorgaSP21} optimize theoretical metrics only, such as average link capacity \cite{DBLP:conf/icc/PiRWZZL22} and estimated hop count \cite{DBLP:conf/networking/LinLLLF22}. Their obliviousness to the practical traffic flows could lead to overcrowded or idle links, eventually affecting the overall network performance.   
Furthermore, they consider a fixed power allocation scheme for satellite laser terminals, without the concern of emerging space lasers \cite{mynaric}, which could be equipped with fine-grained power adjustment capability to further reduce energy consumption.

To fill the research gap and tackle the aforementioned challenges, we propose \textsc{SatFlow}, a scalable network planning framework for LEO mega-constellations. 
Our framework aims at optimizing the total operational costs for satellites’ terminal communication and ISL re-establishment. Specifically, 
considering that network topology, traffic regulation, and link capacities could all affect the costs, 
we optimize (1) the ISL re-establishment scheme for dynamic network topology, (2) the traffic allocation on each ISLs, and (3) the allocated power for satellite laser terminals.  
Since traffic and power allocation affect the energy cost only, and they could be optimized based on a given topology, we decompose our problem hierarchically into two sub-problems and propose two-tier modules to tackle them. Specifically, the upper-level module deals with the ISL re-establishment sub-problem to optimize the total costs, while the lower-level module tackles the traffic and power allocation sub-problem to minimize the energy consumption under a fixed network topology. We propose a temporal graph learning-assisted multi-agent reinforcement learning (RL) algorithm for the upper-level problem and a distributed Lagrangian-based optimization algorithm for the lower-level one.

Our main contributions are summarized as follows:
\begin{itemize}
    
    \item We formulate a novel network planning optimization problem to jointly plan for traffic allocation, satellite terminal power allocation, and ISL re-establishment pattern, in large-scale mega-constellations. 
    
    \item We present the first work to consider power fine-tuning strategies for satellite terminals to optimize energy saving in mega-constellations with dynamic traffics.
    
    \item Our proposed distributed and hierarchical algorithms capture the underlying dynamics among traffic flows and satellite movement, and are highly scalable to practical mega-constellation sizes.
    
    \item Extensive simulations on mega-constellations show that \textit{SatFlow} could reduce the violation of flow conservation by up to $21.0\%$ and reduce the total costs by up to $89.4\%$, compared with various state-of-the-art benchmarks.
\end{itemize}

The remainder of the paper is organized as follows.  The literature review on network planning is conducted in Section \ref{sec:literature} first. Section \ref{sec:formulation} then presents the modeling of our system and the formulation of our problem. Section \ref{method} elaborates on our solution to the problem. Performance evaluation is given in Section \ref{sec:performance}. We draw the conclusion in Section \ref{sec:conc}.

\begin{figure}[t]
    \centering
    \includegraphics[scale=0.46]{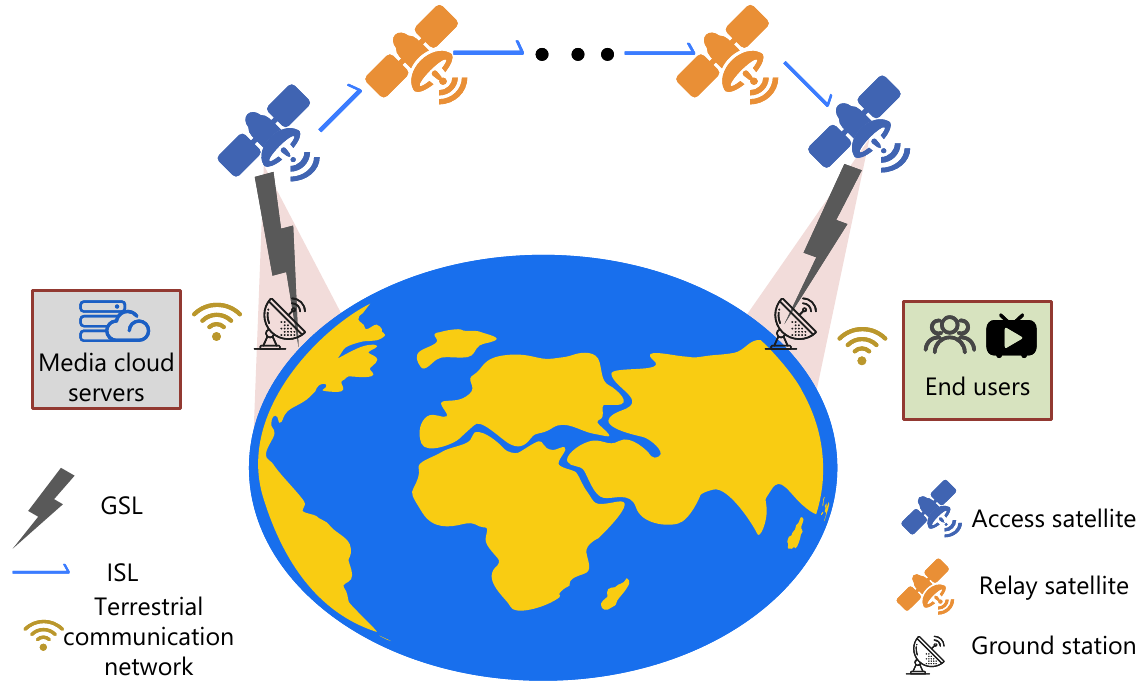}
    \caption{Data transmission in ISL-enabled LEO constellations.}
    \label{fig:scenario}
\end{figure}

\section{Related Work}
\label{sec:literature}

\subsection{Terrestrial Network Planning for Cost Optimization}
Existing works on terrestrial network planning are dedicated to determining the traffic capacity of edges (e.g. fiber links) or nodes (e.g. base stations) \cite{DBLP:conf/sigcomm/ZhuGATZJ21,DBLP:conf/sigcomm/AhujaGDBGZLXZ21, DBLP:conf/noms/PremsankarGFV18}, and/or network topology \cite{DBLP:conf/infocom/ZhaoWWW14,DBLP:conf/infocom/WangH16,DBLP:conf/sigcomm/ZhuGATZJ21,DBLP:conf/sigcomm/AhujaGDBGZLXZ21,DBLP:conf/noms/PremsankarGFV18} to reduce  deployment costs, under the traffic demand or QoS constraints. For example, the number of base stations and their connection status with the user equipment in cellular networks have been optimized using approximation algorithms, in order to minimize the total deployment cost of the base stations under the traffic demand and latency  constraints\cite{DBLP:conf/infocom/WangH16}. Zhu \textit{et al.} \cite{DBLP:conf/sigcomm/ZhuGATZJ21} use a combination of RL and optimization solvers to determine the amount of optical fiber links and IP capacity of a given network, in order to minimize fiber procurement and wavelength provision costs.
However, all these planning techniques could not generate a sustainable and satisfying plan in a reasonable time for large-scale mega-constellations. It is either due to the explosive action space in centralized RL \cite{DBLP:conf/sigcomm/ZhuGATZJ21} or the large variable size when using traditional solvers   \cite{DBLP:conf/sigcomm/AhujaGDBGZLXZ21,DBLP:conf/noms/PremsankarGFV18}. Our work decomposes the original problem into two-tier hierarchical sub-problems which are further solved by carefully-crafted distributed algorithms with great scalability and parallelism capability.

\subsection{Network Planning in LEO Mega-constellations} 
Network planning in existing LEO mega-constellation studies mostly focus on  ISL re-establishment \cite{DBLP:conf/icc/PiRWZZL22,DBLP:conf/networking/LinLLLF22,DBLP:journals/twc/MayorgaSP21,DBLP:conf/globecom/LiLRP23} and resource allocation \cite{DBLP:journals/twc/MayorgaSP21,DBLP:conf/globecom/LiLRP23}.
 Researchers in \cite{DBLP:conf/icc/PiRWZZL22,DBLP:conf/networking/LinLLLF22}  adjust the ISL linkage scheme to improve the throughput and lower the ISL establishment or switching cost. Mayorga \textit{et al.} \cite{DBLP:journals/twc/MayorgaSP21} further consider network resource (e.g. multiple access like CDMA) allocations apart from ISL matching, to maximize the average transmission data rate. 
Furthermore, LEO mega-constellations have been further explored to cooperate with terrestrial segments (e.g. base stations) to jointly optimize network performance \cite{DBLP:conf/infocom/LaiLWLXW22} or maintenance cost \cite{DBLP:conf/infocom/LaiLWWDLLW23,DBLP:conf/icnp/LaiLZWW21}. 
However, these works neither take account of traffic dynamics, nor do they consider fine-tuning the satellite terminal power for customized channel capacity to save energy. Our work, on the other hand, tackles the practical network planning problem with traffic dynamics and fully exploits the potential of time-variant terminal power adjustment in energy saving.

\section{System model and problem formulation}
\label{sec:formulation}

\subsection{LEO Mega-constellation Architecture and Modeling}
\label{leo}

We consider data transmission between global ground stations via LEO mega-constellations. Take multimedia transmission as an example, as illustrated in Fig \ref{fig:scenario}, multimedia content (e.g., videos) is initially sent from media cloud servers to a ground station via terrestrial networks like 5G or Wi-Fi. The content is then uploaded to a nearby LEO access satellite through a ground-satellite link (GSL) and relayed via ISLs to a ground station near the end users. The videos are then downloaded by the users through a terrestrial network. The network operator needs to (1) select ISL paths for all flows, (2) adjust satellite terminal power for varying data rates, and (3) determine the constellation topology by re-establishing ISLs.

Consider $M$ LEO satellites are evenly distributed in $N_{orbit}$ orbital planes at altitude $h$ and inclination $\epsilon$. Satellites are indexed from 0 to $M-1$ orbit by orbit, forming a set $\mathcal{M}$.
Each satellite $m\in\mathcal{M}$ is equipped with four laser terminals for establishing four possible full-duplex ISLs (one for each terminal). 
Following the typical ISL architecture \cite{DBLP:conf/infocom/LyuWLLLL23}\cite{DBLP:conf/infocom/CaoZ23}, each satellite connects with neighboring satellites in the same orbit through two permanent intra-plane ISLs, and links with satellites in adjacent orbits through two inter-plane ISLs, one eastward and one westward. 
  Additionally, satellites can only be connected via an ISL if their distance is below a threshold for reliable transmission\cite{DBLP:conf/icc/PiRWZZL22,DBLP:conf/globecom/LiLRP23}.

 We define the adjacency matrix $A^t$, whose entry $a_{u,v}^t$ is 1 when an ISL is established from satellite $u$ to $v$ at time $t$, and 0 otherwise. $\mathcal{A}^t$ represents the set of adjacency matrices meeting the above ISL establishment constraints.
We assume the terminal power is adjusted every $D_e$ seconds, forming the time-slot set $\mathcal{T}_e=\{t_0, t_1, ..., t_{n-1}\}$. On the other hand, ISLs could be re-established every $D_i$ seconds. We assume $D_i\gg D_e$ 
due to the wear and tear on mechanical arms if performing frequent re-establishment.
For simplicity, $D_i$ is set as $k$ ($k\in\mathcal{Z}$, $k>1$) multiples of $D_e$, and the centralized time slots of conducting ISL re-establishment form a set $\mathcal{T}_i=\{t_0,t_k,...,t_{\lfloor (n-1)/k \rfloor k\}}$.

\subsection{Communication Model}
\label{comm}
When data transmits from satellite $m$ to $n$ through an established ISL in space using laser beams at time $t$, it experiences a free-space path loss $L_{m,n}^t$, calculated \cite{fs} as: 

\begin{equation}
\label{eq:loss}
    L_{m,n}^t=(\frac{4\pi d_{m,n}^t f}{c})^2
\end{equation}
where $d_{m,n}^t$ is the Euclidean distance of satellite $m$ and $n$ at time $t$, $f$ is the carrier frequency, and $c$ is the light speed.

Then, according to the Shannon theorem and the modeling of inter-satellite free-space signal-to-noise ratio  \cite{DBLP:conf/icc/PiRWZZL22}, the channel capacity $C_{m,n}^t$ from satellite $m$ to $n$ is:

\begin{equation}
\label{eq:shannon}
    C_{m,n}^t = B\log_2(1+\frac{p_{m}^tG_mG_n}{k_B\tau BL_{m,n}^t})
\end{equation}
where $B$ is the channel bandwidth,
$p_{m}^t$ is the allocated power of satellite $m$ for establishing the ISL at time $t$. $G_m$ and $G_n$ are the antenna gain of the satellite $m$ and $n$, respectively. $k_B$ is the Boltzmann constant, and $\tau$ is the thermal noise.

Furthermore, established ISLs require a minimum data rate threshold for transmission, denoted as $C_{min}$, to complete necessary handshakes \cite{DBLP:journals/twc/MayorgaSP21}. Combining Eq. \eqref{eq:loss} and \eqref{eq:shannon}, the required power  $\mathcal{P}_{m,n}^t$ allocated for transmission with a data rate of $C$ from satellite $m$ to $n$ at time $t$ would be:

\begin{equation}
\label{eq:Q}
     \mathcal{P}_{m,n}^t(C) = a_{m,n}^t\frac{k_B\tau B}{G_m G_n}(\frac{4\pi d_{m,n}^t f}{c})^2(2^{\frac{\max(C_{min},C)}{B}}-1)
\end{equation}

\subsection{ISL Switching Cost Model}
\label{swi}

      We model the operational cost during ISL re-establishment, primarily from laser terminal rotation, termed ISL switching cost. It could be estimated to be proportional to the rotation angle of laser terminals \cite{DBLP:journals/taes/LeeC21}. We assume satellite $m$ links with satellite $v_{m(e)}^t$ and $v_{m(w)}^t$ at its eastern and western neighboring orbit at time $t$, respectively. To prevent complete communication interruption, we use an asynchronous scheme ensuring some ISLs remain connected at any time. 
  Assume during $[t-\eta,t+\eta]$, ISLs are allowed to be re-established, where $t\in\mathcal{T}_i$, and $2\eta$ is a short period required for all ISLs to conduct re-establishment. 
  The total rotation angle  $\Theta_{m}^{t}$ induced by satellite $m$'s laser terminals during  $[t-\eta,t+\eta]$ could then be estimated as the sum of the rotation angle of its eastern and western laser terminals at the static time $t$:

\begin{equation}
\label{eq:angle}
    \Theta_{m}^{t} = \theta_{m,v_{m(e)}^{t-\eta},v_{m(e)}^{t+\eta}}^{t} + \theta_{m,v_{m(w)}^{t-\eta},v_{m(w)}^{t+\eta}}^{t}
\end{equation}
where $\theta_{u,v_1,v_2}^t$ represents the rotation angle of satellite $u$'s ISL laser terminal from aligning satellite $v_1$ to $v_2$ at time $t$.
Finally, the total ISL switching cost $S_t$ during  $[t-\eta,t+\eta]$ is proportional to the sum of $\Theta_{m}^{t}$ for all $m\in\mathcal{M}$, calculated as:

\begin{equation}
\label{eq: sw}
    S_t=\mu\sum_{m\in\mathcal{M}}\Theta_{m}^{t}
\end{equation}
where $\mu$ is the unit switching cost per adjusted steering angle.

\subsection{Traffic Flow Model}
\label{flow}

  We assume traffic flows are designated to transmit through a shell of LEO mega-constellations, and their meta-information is pre-estimated for each period between adjacent time slots in $\mathcal{T}_i$. Suppose for a period $\mathcal{T}$, $n_t  (t\in\mathcal{T})$ aggregated flows are estimated to transmit,  denoted as $\Omega_t=\{\omega_t^1, \omega_t^2, ..., \omega_t^{n_t}\}$.
 To reduce the end-to-end latency, we consider the flow $\omega\in\Omega_t$ to be transmitted from the satellite (denoted as $src_{\omega}$) closest to the source ground station to the satellite (denoted as $dst_{\omega}$) closest to the destination through ISLs only, with a requested data rate $d_{\omega}$. 
For ease of flow regulation, we enforce that when the constellation topology is unchanged, the traffic allocation on all ISLs remain fixed. Furthermore, for better link utilization, we allow for splittable traffic allocation where flows could be split and allocated on different paths, which is broadly used in modern wireless communication networks for energy saving \cite{DBLP:journals/tmc/TangZY19,DBLP:conf/icc/TaiBHA06,DBLP:conf/infocom/JiaLD04}.

\subsection{Problem Formulation}
\label{prob}

Our goal is to minimize the total operational cost including the satellites' energy consumption for transmission and the ISL switching cost, in the analyzed period. We optimize the ISL re-establishment plan $A^t$ for next $D_i$ at each $t$ when $t+\eta\in\mathcal{T}_i$, and decide the allocated data rate $Y_{m,n}^{\omega,t}$ for flow $\omega\in\Omega_t$ from satellite $m$ to $n$ (i.e. the traffic allocation) for each  $t\in\mathcal{T}_e$. The allocated data rate further guides the laser terminals to adjust the power allocated for established ISLs according to Eq. \eqref{eq:Q}. Note that in this work, we have set a fixed asynchronous scheduling scheme for ISL re-establishment between two topologies, thus leaving $\eta$ in Sec.\ref{swi} as a hyper-parameter. It also means that at time $s$ between $[t-\eta,t+\eta]$ for $t\in\mathcal{T}_i$, $A^s$ could be determined once the ISL re-establishment plan is generated in $t-\eta$. We do not optimize the ISL re-establishment scheduling during $[t-\eta,t+\eta]$ since $\eta$ is assumed much smaller than $D_i$, and thus the scheduling does not affect our optimization too much. Instead, it only provides a protocol avoiding a complete link interruption.

For brevity of problem formulation, we introduce a column vector \textbf{b$^\omega$} for flow $\omega$, with $[\textbf{b}^\omega]_a$ denoting its $(a+1)^{th}$ row:

\begin{align}
\begin{split}
[\textbf{b}^\omega]_a= \left \{
\begin{array}{ll}
    d_\omega,                    & a=src_\omega\\
    -d_\omega,     & a=dst_\omega\\
    0,                                 & otherwise
\end{array}
\right.
\end{split}
\end{align}

The objective and constraints could then be expressed as:

\begin{equation}
\begin{aligned}
\label{problem}
 {\min_{A,Y}}  \quad  &\alpha D_e\sum_{t\in\mathcal{T}_e}\sum_{m,n\in\mathcal{M}} \mathcal{P}_{m,n}^t(\sum_{\omega\in\Omega_t}Y_{m,n}^{\omega,t})+\beta\sum_{t\in\mathcal{T}_i}S_t
 \end{aligned}
\end{equation}
\vspace{-0.5cm}

\begin{align}
\label{c1} 
 s.t. \quad &\mathcal{P}_{m,n}^t(\sum_{\omega\in\Omega_t}Y_{m,n}^{\omega,t})\le a_{m,n}^{t} P_{max}  \quad\forall{m,n\in\mathcal{M}}, \forall{t\in\mathcal{T}_e}\\
\label{c2}
& \sum_{n\in\mathcal{M}}Y_{m,n}^{\omega,t}-\sum_{n\in\mathcal{M}}Y_{n, m}^{\omega,t}=[\textbf{b}^\omega]_m \nonumber\\
& \quad\quad\quad\quad\quad\quad\quad \quad  \quad \forall{\omega\in\Omega_t}, \forall{m\in\mathcal{M}}, \forall{t\in\mathcal{T}_e}\\
\label{c6}
& 0\le Y_{m,n}^{\omega,t} \le a_{m,n}^{t} d_{\omega}  \nonumber\\
& \quad\quad\quad\quad\quad\quad\quad  \forall{\omega\in\Omega_t},\forall{m,n\in\mathcal{M}}, \forall{t\in\mathcal{T}_e}\\
\label{c7}
& A^t\in\mathcal{A}^t \quad\quad\quad\quad\quad\quad\quad\quad\quad\quad\quad\quad \forall{t\in\mathcal{T}_e}
\end{align}
where $\alpha$ and $\beta$ are the positive weights reflecting the preference between terminal energy consumption and ISL switching costs. 
Constraint (\ref{c1}) sets the terminal power limit $P_{max}$. Constraint (\ref{c2}) is the flow conservation constraint for successful flow transmission. Constraint (\ref{c6}) ensures that the amount of traffic flows routed on each ISL should be non-negative but not exceed the flow demand, and no traffic flows are allocated on unestablished ISLs. Constraint (\ref{c7}) ensures that the adjacency matrix should satisfy the constraints in Sec. \ref{leo}.

Problem \eqref{problem} is a mixed-integer nonlinear programming problem, which is tedious to solve. The Fixed Charge Network Flow problem, which proves \cite{DBLP:journals/networks/HochbaumS89} to be NP-hard, could be reduced to it. Therefore, problem \eqref{problem} is also NP-hard. 
We omit the detailed proof due to space limitations. Consequently, the exact solution could not be obtained in an affordable time under the scenario of mega-constellations with a large variable size, motivating us to develop a satisfactory scalable solution.

\begin{figure}[t]
    \centering
    \includegraphics[scale=0.24]{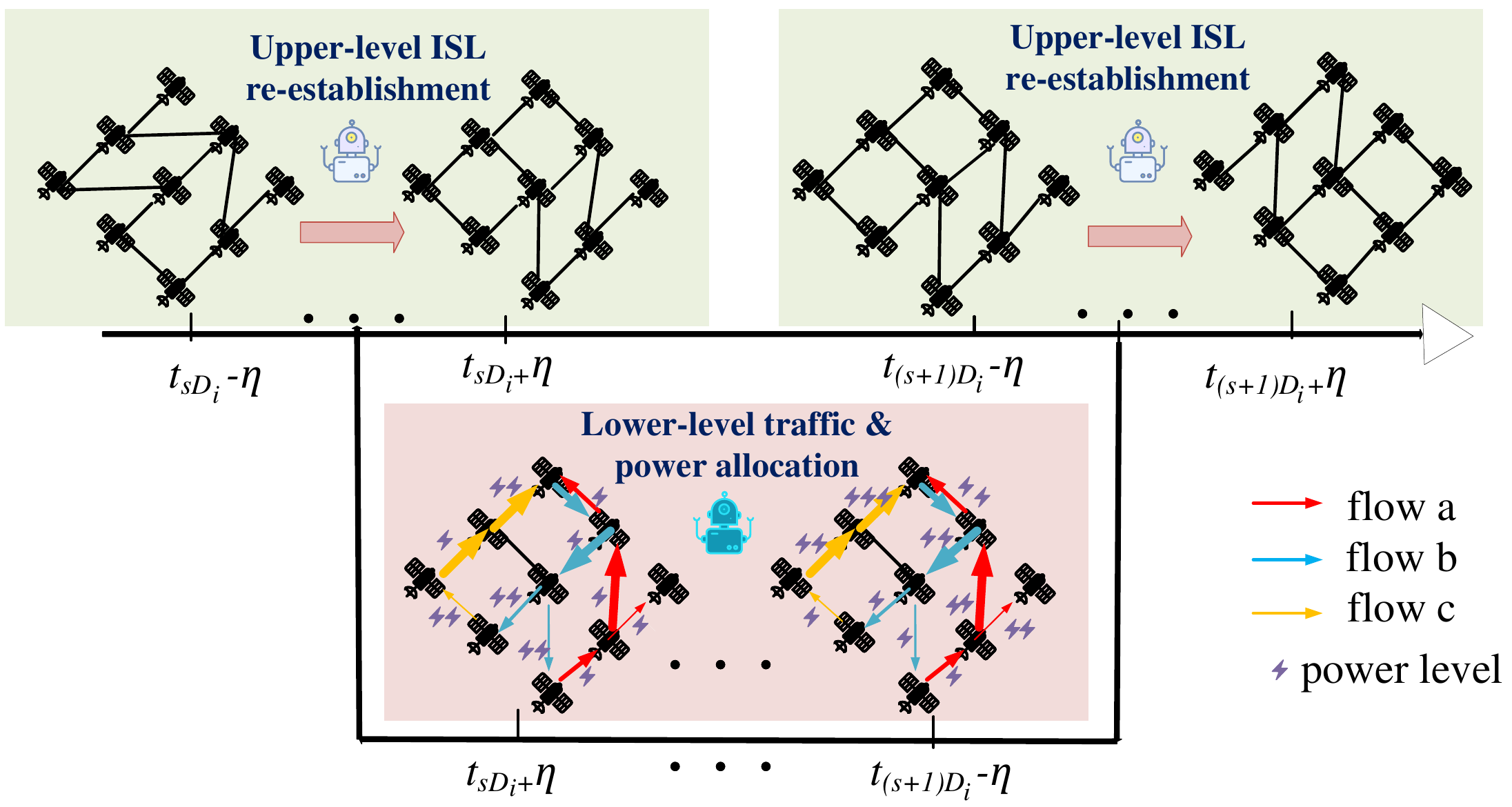}
    \caption{Hierarchical planning framework after the $(s+1)^{th}$ decision for ISL re-establishment. 
   Line boldness for a flow and the number of lightning symbols on an ISL reflect the quantity of reserved data rates and the allocated power, respectively.}
    \label{fig:sat2}
\end{figure}

\section{Algorithm Design}
\label{method}

Digging into the relationship of variables in problem \eqref{problem}, we observe that the traffic allocation scheme affects the energy cost only and could be optimized based on a given topology, while the ISL re-establishment scheme affects the total cost in the long run.  
Therefore, problem \eqref{problem} could be decomposed into two-tier hierarchical sub-problems, illustrated in Fig. \ref{fig:sat2}. To be specific, the lower-level module, elaborated in Sec. \ref{sec:lower}, optimizes the traffic allocation and dynamically allocates the power to save terminal energy on a given topology and traffic. On the other hand, the upper-level module, elaborated in Sec. \ref{sec:upper}, conducts an ISL re-establishment scheme every $D_i$ to lower both the terminal energy consumption and the ISL switching cost. The upper-level module makes an iterative search with the help of our lower-level module, which calculates the energy cost for each given topology. Consequently, the hierarchical decomposition simplifies the problem by decoupling the decision variables efficiently.
To tackle the scalability issue, distributed algorithms based on distributed Lagrangian optimization and multi-agent RL, are proposed for our lower-level and upper-level modules, respectively.

\subsection{Lower-level Traffic and Power Allocation Module}
\label{sec:lower}

Given the ISL re-establishment scheme proposed by the upper-level module, our lower-level module is run for a period under the current fixed topology consisting of all ISLs not experiencing ISL re-establishment. Consider the lower-level module aims to minimize the overall energy consumed by satellite terminals during a period $\mathcal{T}$.
Since the allocated traffic is constrained to be unchanged for the period, we discard the superscript $t$ of $Y_{m,n}^{\omega,t}$ for brevity, reflecting the allocated data rate of flow $\omega$ from satellite $m$ to $n$ during the whole $\mathcal{T}$. We further denote $\mathcal{N}_m$ as the index list of satellites directly connecting with satellite $m$ , then the traffic and power allocation sub-problem during $\mathcal{T}$ could be expressed as:

\begin{equation}
\begin{aligned}
\label{problem_r1}
 {\min_{Y}}  \quad  &\sum_{t\in\mathcal{T}}\sum_{m\in\mathcal{M}}\sum_{n\in\mathcal{N}_m} \mathcal{P}_{m,n}^t(\sum_{\omega\in\Omega_t}Y_{m,n}^{\omega})
 \end{aligned}
\end{equation}

\begin{align}
\label{c1_a} 
 s.t. & \quad\eqref{c1}-\eqref{c6}  \nonumber
\end{align}

\begin{algorithm}[t]
\caption{Lower-level Distributed Traffic and Power Allocation Algorithm  (\textit{SatFlow-L})}
\begin{algorithmic}[1]
\label{alg:low}
\REQUIRE 
the adjacency matrix $A^t$ and the estimated flow collection $\Omega_t$ for $t\in\mathcal{T}$; the sub-gradient iteration limit $k_{sg}$; the training episode limit $k_{max}$; convergence tolerance $\epsilon$.\\
\ENSURE 
allocated traffic $\textbf{Y}$, and terminal power allocation $\mathcal{P}$. \\
\STATE Initialize $\textbf{Y}^{(0)}$ and \textbf{$\lambda^{(0)}$} of iteration 0.\;\\
\FOR{$k=0$ to  $k_{max}$}
\STATE \texttt{// Distributed implementation}
\FOR{$m\in\mathcal{M}$}
\STATE Obtain $\hat{\textbf{Y}}_m^{(k)}$ by minimizing \eqref{eq:inner} onto $\mathcal{Y}$, using projected sub-gradient algorithm up to $k_{sg}$ iterations.

\ENDFOR
\FOR{$m\in\mathcal{M}$}
\STATE Update $\lambda_m^{(k+1)}$ by Eq. \eqref{eq:Lag} and $\textbf{Y}_m^{(k+1)}$ by Eq. \eqref{eq:primal}.\\
\STATE Allocate power $\mathcal{P}_{m,n}^t$ using Eq.  \eqref{eq:Q} for each $n\in\mathcal{N}_m^t$.\\
\ENDFOR
\STATE Break if $\sum_{m\in\mathcal{M}}|\textbf{Y}_m^{(k+1)}-\textbf{Y}_m^{(k)}|\le\epsilon$.
\ENDFOR

\end{algorithmic}
\end{algorithm}

Since Sub-problem \eqref{problem_r1} is a constrained convex optimization problem (the proof is omitted here), centralized Lagrangian optimization is a solution. However, it could induce a slow convergence speed and huge inter-satellite communication costs due to a large variable size.
We thus propose a distributed traffic and power allocation algorithm for our lower-level module, referred to as \textit{SatFlow-L}.

To decompose Sub-problem \eqref{problem_r1} for distributed optimization, we first transform it into an extended monotropic optimization problem \cite{monotropic}, which aims to optimize a sum of convex functions consisting of disjoint sets of variables, subject to equality constraints for a summation of the sets of variables undergone by a certain affine transformation. To adapt our flow conservation constraints to the extended monotropic optimization, we design a linkage matrix $[L_m]_{M\times 4}$ for satellite $m$, which reflects $m$'s neighbours within the whole constellation. Its $(i+1)^{th}$ row and $(j+1)^{th}$ column is:

\begin{align}
\begin{split}
[L_m]_{i,j}= \left \{
\begin{array}{ll}
    1,                    & i=m\\
    -1,     & i\in\mathcal{N}_m,  j:\mathcal{N}_m[j]=i\\
    0,                                 & otherwise 
\end{array}
\right.
\end{split}
\end{align}

Then our extended monotropic optimization problem is:

\begin{equation}
\begin{aligned}
\label{problem_r2}
 {\min_{Y}}  \quad  &\sum_{m\in\mathcal{M}}(\sum_{t\in\mathcal{T}^s}\sum_{n\in\mathcal{N}_m} \mathcal{P}_{m,n}^t(\sum_{\omega\in\Omega_t}Y_{m,n}^{\omega}))
 \end{aligned}
\end{equation}

\vspace{-0.6cm}

\begin{align}
\label{c1_r} 
 s.t. \quad & \sum_{m\in\mathcal{M}}L_m \textbf{Y}_m^\omega = \textbf{b}^\omega \quad\quad\forall{\omega\in\Omega_t}\\
\label{c2_r}
& \textbf{Y}_m\in\mathcal{Y}_m \quad\quad\quad\quad\quad\quad\forall{m\in\mathcal{M}}
\end{align}
where \textbf{Y$_m^\omega$} = $[Y_{m,\mathcal{N}_m[0]}^\omega, Y_{m,\mathcal{N}_m[1]}^\omega,Y_{m,\mathcal{N}_m[2]}^\omega,Y_{m,\mathcal{N}_m[3]}^\omega]^T$,  $\textbf{Y}_m=[\textbf{Y}_m^{\omega_t^1}, \textbf{Y}_m^{\omega_t^2},...,\textbf{Y}_m^{\omega_t^{n_t}}]$, and $\mathcal{Y}_m$ is the set of $\textbf{Y}_m$ comforting Constraint \eqref{c1} and \eqref{c6}. Constraint \eqref{c1_r} corresponds with the flow conservation constraint \eqref{c2}.

 Our \textit{SatFlow-L} algorithm is based on the distributed alternating step method \cite{JE}, which decomposes the original Lagrangian function for an extended monotropic optimization problem into smaller independent sub-functions, achieving much faster optimization speed under parallel computation. We show the pseudo-code of \textit{SatFlow-L} in Algorithm \ref{alg:low}. To be specific, each satellite $m$ is treated as a distributed unit, with its local Lagrangian function $\mathcal{L}_m(\hat{\textbf{Y}}_m, \textbf{Y}^{(k)}, \lambda^{(k)})$ at the $k^{th}$ iteration expressed as:

\begin{equation}
\label{eq:inner}
\begin{aligned}
&\sum\limits_{t\in\mathcal{T}}\sum\limits_{n\in\mathcal{N}_m} {\mathcal{P}}_{m,n}^t(\sum_{\omega\in\Omega_t}\hat{Y}_{m,n}^\omega)+\sum\limits_{\omega\in\Omega_t}(\left<\lambda^{\omega(k)}, L_m \hat{\textbf{Y}}_m^{\omega(k)}\right>+\frac{\rho}{2}\\
    &\sum\limits_{l=1}^M([L_m]_l \hat{\textbf{Y}}_m^{\omega(k)}-[L_m]_l \textbf{Y}_m^{\omega(k)} + \frac{1}{q_l}(\sum\limits_{j=1}^M [L_j]_l \textbf{Y}_j^{\omega(k)} -  [b^\omega]_l))^2)
\end{aligned}
\end{equation}
where $\left<\cdot,\cdot \right>$ is the inner product, $\mathbf{\lambda}$ is an $M$ by $n_t$ matrix with the entry $\lambda_m^\omega$ denotes the Lagrangian multiplier for satellite $m$ and flow $\omega$, $\rho$ is the penalty coefficient for the flow conservation violation, and $q_l$ is the number of linkage matrices with a nonzero $l^{th}$ row.
For the $(k+1)^{th}$ iteration, each satellite minimizes  \eqref{eq:inner} with regard to $\hat{\textbf{Y}}_m$ in parallel, conducted by several steps of gradient descent projected on the sets \eqref{c2_r} up to $k_{sg}$ iterations for timing concerns. Then, $\lambda_m^{(k+1)}$ and $\textbf{Y}^{(k+1)}$ are updated distributively as:

\begin{equation}
\label{eq:Lag}
    \lambda_m^{(k+1)}=\lambda_m^{(k)}+\frac{\rho \sigma}{q_m}(\sum\limits_{i=1}^M [L_i \hat{\textbf{Y}}_i^{(k)}]_m - [\textbf{b}]_m)
\end{equation}

\begin{equation}
\label{eq:primal}
    \textbf{Y}_m^{(k+1)}=\textbf{Y}_m^{(k)}+\sigma(\hat{\textbf{Y}}_m^{(k)}-\textbf{Y}_m^{(k)})
\end{equation}
where $\sigma\in[0,2]$ is the step size for update. Armijo line search \cite{arm} is employed to determine the step size for gradient descent, with the benefit of robustness to non-convex functions and fast convergence.

\begin{theorem}
\label{theorem2}
    \textit{SatFlow-L} converges to the optimal solution $Y^{*}$ of Sub-problem \eqref{problem_r1}.
\end{theorem}

We could prove Theorem \ref{theorem2} through the following steps: 1) the converted problem \eqref{problem_r2}$\sim$\eqref{c2_r} is an extended convex monotropic optimization problem; 2) our \textit{SatFlow-L} algorithm converges \cite{JE} to optimal for the converted problem; 3) the converted problem is equivalent to Sub-problem \eqref{problem_r1}. The detailed proof is omitted due to space limitations.

We further analyze the communicational and computational efficiency of \textit{SatFlow-L} in the following theorems.

\begin{theorem}
\label{theorem3}
   The distributed implementation of each satellite $m$ in \textit{SatFlow-L} requires the following variables from satellite $i\in\mathcal{M}$: $\lambda_i^{(k)}$ of $i$ within one-hop only and $\textbf{Y}_i^{(k)}$ of $i$ within two-hop only.
\end{theorem}

\begin{theorem}
\label{theorem5}
  The time complexity of \textit{SatFlow-L} is $O(n_t)$ for $t\in\mathcal{T}$.
\end{theorem}

The proof of the above theorems could be derived by mathematically analyzing Line 5 and 8 of Algorithm \ref{alg:low}. We omit the proof here due to space limitations. Theorem \ref{theorem3} guarantees that each satellite only needs to exchange variables with its neighbors at most two-hop away. Theorem \ref{theorem5}  reveals that \textit{SatFlow-L} has scalable time complexity which is only linear to the number of existing flows $n_t$, and is independent of the constellation size.

\subsection{Upper-level ISL Re-establishment Module}
\label{sec:upper}

The upper-level module optimizes long-term ISL re-establishment plans within $t\in\mathcal{T}_i$ for total cost optimization. It employs reinforcement learning, a data-driven method for sequential decision-making.
 However, given the vast number of candidate plans in mega-constellations, single-agent RL faces potential convergence issues due to an expansive action space. We address this issue by using multi-agent RL, where each agent optimizes a subset of the action space. We now elaborate on our upper-level multi-agent-RL-based ISL re-establishment algorithm, referred to as $\textit{SatFlow-U}$.

\textbf{Granularity design of RL agents.} Before defining each agent's responsibility, we observe that since each terminal could only establish an ISL as specified in Sec. \ref{leo}, the linkage conflict would occur if a terminal is designated to establish ISLs with multiple satellites. To tackle the issue, we first partition each orbit into $N_G$ groups of adjacent satellites. Each group covers $\lfloor N_{spo}/N_G\rfloor$ satellite(s), where $N_{spo}$ is the number of satellites per orbit. Then, we define the  \textit{index offset} between satellite $m$ and $n$ ($m>n$) as $(m - n)\%n_{spo}$. We restrict that satellites in the same group (managed by an agent) should link eastern-orbit satellites of the same index offset.
We then resolve conflicts by eliminating boundary ISLs.
The group-based joint action selection scheme could resolve linkage conflict without eliminating many ISLs while maintaining numerous re-establishment options.

\textbf{Asynchronous ISL re-establishment scheme.} We now elaborate on our proposed asynchronous ISL re-establishment scheme to prevent a complete link interruption. We set the time window of ISL re-establishment $2\eta$ as $N_Gt_{\theta_{max}}$, where $t_{\theta_{max}}$ the time for the laser terminal to conduct the maximum rotation angle. We enforce ISLs to be re-established group by group. To be specific, during each $t_{\theta_{max}}$, we randomly select a group of satellites in each orbit that has not conducted ISL re-establishment to allow for re-establishing ISLs. Then after $N_Gt_{\theta_{max}}$, all groups of satellites have finished conducting ISL re-establishment. The scheme ensures that at least $N_G-1$ groups of satellites preserve a stable ISL connection at any time, and the ISL re-establishment period is restricted in $2\eta$, which is independent of the number of orbital planes $N_{orbit}$.

\textbf{Architecture and design of our multi-agent RL.} We now elaborate the design of state, action, reward, and training architecture for our multi-agent RL.

\begin{figure}[t]
    \centering
    \includegraphics[width=3.5in,height=3.3in]{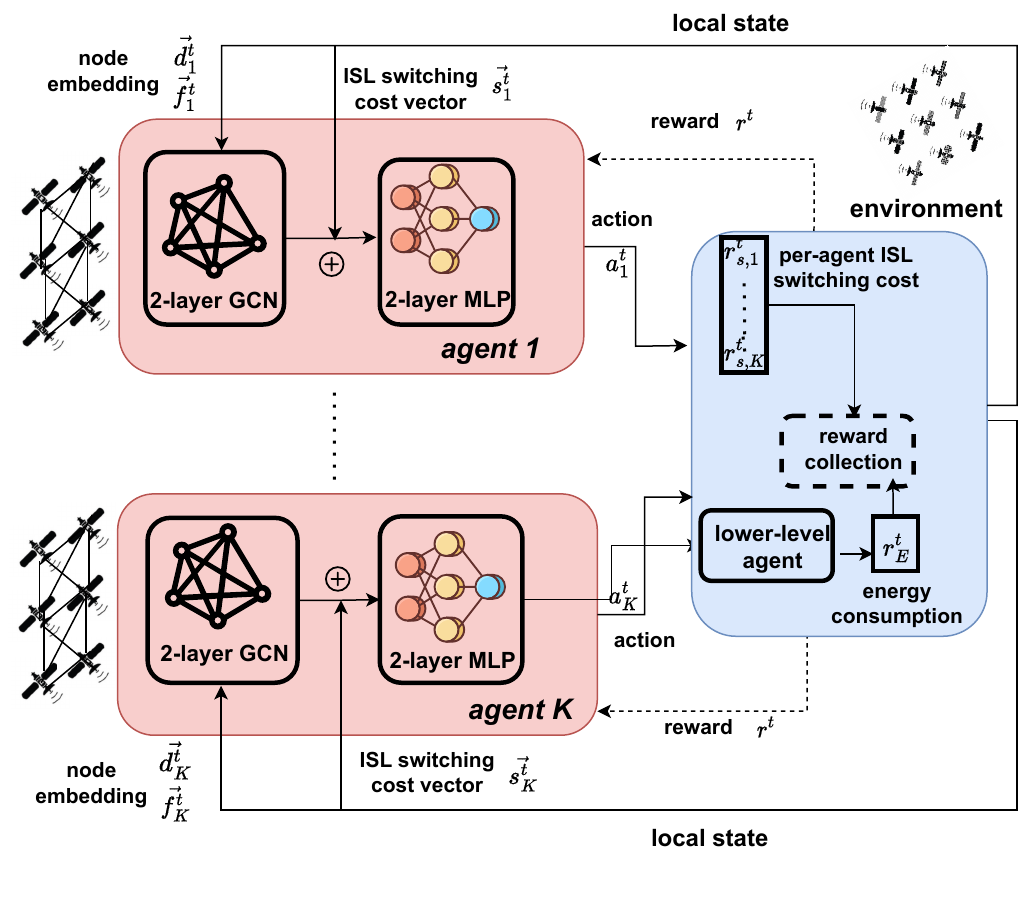}
    \caption{The upper-level module for ISL re-establishment.}
    \label{fig:high_level_agent}
\end{figure}

\textit{1) Feature design and State representation.}
To extract features for an agent to aid the ISL re-establishment selection, we first model the whole constellation as a graph, whose nodes represent all the satellites and edges represent certain relationships between satellite pairs. To adapt to the constellation and traffic dynamics, we embed the temporal information using temporally-weighted normalization \cite{DBLP:conf/www/QuZDS20} into the edge features for prospective concerns. Here are the extract features.

\begin{itemize}
    \item Normalized temporally-weighted maximum data rates on ISL-capable edges, indicating future link cost efficiency.
    
    \item Normalized temporally-weighted flow data rates on edges that connect all pairs of source and destination nodes, reflecting future traffic patterns.

    \item A $\mathcal{W}$-dimensional node feature on all nodes, where $\mathcal{W}$ is the candidate size of index offsets. Its $i^{th}$ entry is the ISL switching cost when choosing the eastern-neighboring satellite of the $i^{th}$ candidate index offset to link with. 
\end{itemize}

To incorporate the first two kinds of edge features into the subsequent graph convolution network (GCN), we transfer them into two node embedding sets using a classical pre-trained graph embedder, 
LINE \cite{DBLP:conf/www/TangQWZYM15}, which can learn node embedding from edge features. After transformation, node $i$'s embedding is denoted as $\Vec{d_{(i)}^t}\in\mathcal{R}^{1\times S}$ and $\Vec{f_{(i)}^t}\in\mathcal{R}^{1\times S}$, corresponding with the two sets of node features respectively, where $S\ll M$ is the embedding dimension. For agent $j$, the resulting embedding of its belonging satellites and their neighbors in the eastern orbit are concatenated to get agent-wise embedding  $\Vec{d_j^t}\in\mathcal{R}^{N_s\times S}$ and $\Vec{f_j^t}\in\mathcal{R}^{N_s\times S}$, where $N_s=2\lfloor N_{spo}/N_G\rfloor$ is the number of the involved nodes.

For the third kind of feature, since we restrict satellites in an agent to choose the same index offset, we aggregate and sum up the feature of all its belonging satellites and get a single $\mathcal{W}$-length vector $\Vec{s_j^t}\in\mathcal{R}^{1\times \mathcal{W}}$, reflecting agent $j$'s induced ISL switching cost for each candidate index offset at time $t$.

Finally, agent $j$'s state is represented as ($\Vec{d_j^t}$, $\Vec{f_j^t}$, $\Vec{s_j^t}$), reflecting future ISL cost efficiency, traffic patterns, and the ISL switching cost of choosing various index offsets, respectively.

\begin{algorithm}[t]
\caption{Upper-level Multi-agent-RL-based ISL Re-establishment Algorithm (\textit{SatFlow-U})}
\begin{algorithmic}[1]
\label{alg:high}
\REQUIRE 
the estimated flow collection $\Omega_t$; maximal training episode $K_{max}$;  the update period of target Q function $k_{update}$; batch size $N_{batch}$.\\
\ENSURE 
a well-trained ISL linkage plan for all $t\in\mathcal{T}_i$. \\

\STATE Initialize primary Q function $Q(\cdot;\theta)$ and target Q function $Q(\cdot;\theta')$, where $\theta=\theta'$; 
\FOR{$k=1$ to  $K_{max}$}

\FOR{$t$ when $t+\eta \in \mathcal{T}_i$}
\FOR{each agent $j$}
\STATE Observe state $s^{j}(t)$, and take action $a^{j}$ based on primary $Q$ function with $\epsilon$-greedy satisfying $\mathcal{A}^t$.

\STATE Get ISL switching cost $r_{s,j}^{t}$ incurred by agent $j$.

\ENDFOR

\STATE Eliminate conflicting ISL(s), and perform Algorithm \ref{alg:low} on all induced topology with static ISLs during $[t,t+D_i]$ to get overall energy consumption $r_E^t$.

\STATE Obtain reward $r^t$ from  $r_{s,j}^t$ and $r_E^t$, using Eq. \eqref{eq:reward}.

\STATE Update state to $\textbf{s}(t+ D_i)$.
\FOR{each agent $j$}

\STATE Add the tuple ($s^{j}(t)$, $a^{j}$, $r^t$, $s^{j}(t+D_i)$) to the replay buffer.
\ENDFOR

\ENDFOR

\STATE Randomly sample $N_{batch}$ tuples from the replay buffer, and update $\theta$ by minimizing the loss defined in Eq. \eqref{eq:q_loss}.

\IF{$k \% k_{update}==0$}
\STATE $\theta'\leftarrow \theta$
\ENDIF

\ENDFOR

\end{algorithmic}
\end{algorithm}

\textit{2) State Aggregation.}
To bridge the relationship between our downstream task and extracted features, we inject each agent's corresponding graph into a GCN. As shown in Fig. \ref{fig:high_level_agent}, the graph for agent $j$ has the concatenated node embedding $\textbf{X}_j=(\Vec{d_j^t},\Vec{f_j^t})\in \mathcal{R}^{N_s\times 2S}$, and the edges connecting all the node pairs that could establish ISLs. The graph then goes through 2-layer GCNs. Let $\textbf{A}_j\in \{0,1\}^{N_s\times N_s}$ be the adjacency matrix for agent $j$'s graph, and $\textbf{D}_j$ be the degree matrix with its diagonal element $\textbf{D}_{j}[i,i]=\sum_{m}\textbf{A}_j[m]$. We denote the trainable parameters of two graph convolutional layers as $\textbf{W}_G^{(0)}\in \mathcal{R}^{S\times C_0}$ and  $\textbf{W}_G^{(1)}\in \mathcal{R}^{C_0\times C_1}$, respectively (where $C_0$ and $C_1$ are the output node dimension of the two layers), then the resulting node embedding $\mathcal{F}_G$ is:

\begin{equation}
    \mathcal{F}_G(\textbf{X}_j,\textbf{A}_j)=\sigma_R (\widetilde{\textbf{A}_j} \sigma_R (\widetilde{\textbf{A}_j} \textbf{X}_j \textbf{W}_G^{(0)})\textbf{W}_G^{(1)})
\end{equation}
where $\widetilde{\textbf{A}_j}=\textbf{D}_j^{-1/2}(\textbf{I}_{N_s}-\textbf{A}_j)\textbf{D}_j^{-1/2}$ ($\textbf{I}_{N_s}$ is an $N_s$-dimensional identity matrix), and $\sigma_R$ is the ReLU activation function. 
The node embedding $\mathcal{F}_G$ is then flattened and concatenated with the features of ISL switching costs $\Vec{s_j^t}$, producing the flattened features $\mathcal{F}_C\in\mathcal{R}^{1\times(N_s C_1+\mathcal{W})}$. They serve as the aggregated states for the future RL input.

\textit{3) Action.}
The action space for an agent is a discrete set of index offsets between its belonging satellites and those in the eastern neighboring orbit that could establish ISLs. As shown in Fig. \ref{fig:high_level_agent}, after the state aggregation, the concatenated features are fed into 2-layer MLPs to output $\mathcal{W}$ values, corresponding to the Q values for candidate index offsets. Then the output of the action
probability distribution is expressed as 
$(\sigma_R \mathcal{F}_C \textbf{W}_M^{(0)} )\textbf{W}_M^{(1)}$, where $\textbf{W}_M^{(0)}\in\mathcal{R}^{(N_s C_1+\mathcal{W})\times M_0}$ and $\textbf{W}_M^{(1)}\in\mathcal{R}^{M_0\times \mathcal{W}}$ are the trainable weights of the two  MLP layers, and $M_0$ is the hidden unit size of the first MLP layer.

\textit{4) Reward.}
To reflect the objective \eqref{problem}, our reward includes the energy consumed by the laser terminals and the ISL switching cost. For the latter component, we sum up all satellites' induced ISL switching costs and get $r_{s}^t  = \sum_{j\in\mathcal{M}}r_{s,j}^t$, 
where $r_{s,j}^t$ is the ISL switching cost induced by agent $j$. For generability, homogeneous multi-agent training is applied so that all agents share the same reward $r^t$, calculated as:

\begin{equation}
\label{eq:reward}
    r^t = \alpha(1-\widetilde{r_{E}^t})+\beta (1- \widetilde{r_s^t})
\end{equation}
where $\widetilde{r_{E}^t}$ and $\widetilde{r_{s}^t}$ are the normalized energy consumed during $[t, t+D_i]$ and the normalized ISL switching cost during $[t, t+2\eta]$, respectively.

\textit{5) Training Architecture.}
  We adopt deep-Q network learning to optimize the ISL re-establishment pattern every $D_i$. The detailed training pseudo-code is shown in Algorithm \ref{alg:high}. After all the agents take the actions (Line 5), the lower-level module is implemented (Line 8) to help produce the joint reward $r_t$ according to Eq. \eqref{eq:reward} (Line 9), which is fed to each agent for DQN training. Let  $\textbf{s}$ and $\textbf{a}$ represent the state and action, respectively, and trainable $\theta=[\textbf{W}_G^{(0)}, \textbf{W}_G^{(1)}, \textbf{W}_M^{(0)},\textbf{W}_M^{(1)}]$, then the loss $L(\theta)$ to be minimized w.r.t. $\theta$ is defined as:
  \begin{equation}
  \label{eq:q_loss}
      L(\theta)=\sum_{i=1}^{N_{batch}}(r_i+\phi\max_{\textbf{a}}Q(\textbf{s}'_i,\textbf{a};\theta')-Q(\textbf{s}_i,\textbf{a}_i;\theta))
  \end{equation}
  where $(\textbf{s}_i,\textbf{a}_i,r_i,\textbf{s}'_i)$ is the $i^{th}$ sampled tuple, $\theta'$ is the fixed original parameter, and $\phi$ is the discounted factor.

\textbf{\textsc{SatFlow} at runtime.}
For each upper-level decision round, a global coordinator is selected either in the terrestrial command center or in a primary satellite. It gathers satellites' geo-locations and flow meta-information, calculates satellites' embedding, and then pushes the overall topology and satellites' local embedding back to the satellites. 

A group leader from each satellite group, selected following certain selection policies (e.g., random in this paper), takes the embeddings from all satellites within the group and runs the policy network trained by Algorithm \ref{alg:high} to generate the index offset decision. Note that a single DNN-based policy network incurs little overhead, e.g. less than 5 ms for the inference on a Jetson TX2, a typical satellite edge computing testbed \cite{DBLP:journals/tmc/ZhuLZW24,DBLP:conf/asplos/DenbyL20}. This validates the feasibility of a timely inference for the upper-level decision in reality. 
An episode stops once the coordinator finds it reaching convergence and sends back the termination command to each group leader.

For the lower-level traffic and power allocation module, each satellite retrieves the topology from the global coordinator, the Lagrangian variables from its one-hop neighbors, and the variables of allocated data rates from its one- and two-hop neighbors. Then it calculates the updated corresponding variables locally. The lower-level module is terminated once the coordinator finds it reaching convergence and sends back the termination command to each satellite.

\section{Performance Evaluation}
\label{sec:performance}

\begin{table}[t]
    \centering
      \caption{Number of flows for various traffic intensity levels.}
    \begin{tabular}{c|cccc}
    \hline
     Shell $\backslash$ Traffic intensity & Low & Medium & High \\ 

     \hline
        Starlink Shell A   &   [10,20)  &  [20,30)  & [30,40]  \\
        Starlink Shell B     &   [20,40)  &  [40,60)  & [60,80]  \\
        Kuiper  &   [20,30)  &  [30,40)  & [40,50]\\
        \hline
       
    \end{tabular}
  
    \label{tab:flow}
\end{table}

\subsection{Experimental Setup}

\textbf{Constellation setting.} To evaluate the performance of our algorithm in real-world mega-constellations, we use \texttt{PyEphem} \cite{pyephem} to simulate three shells of LEO constellations, two from Starlink \cite{Starlink} (denoted as ``Shell A'' and ``Shell B'') and one from Kuiper \cite{kuiper}. 
They have 4, 6, and 28 orbits each, 43, 58, and 28 satellites per orbit each, with the inclination of 97.6\degree, 97.6\degree, and 33.0\degree each, and the altitude of 560 km, 560 km, and 590 km each. We set the total analyzed period as 7200 s, which is close to the orbiting period of the selected shells. We then set the interval of power adjustment $D_e$ as 30 s, the interval of ISL re-establishment $D_i$ as 1200 s, and the maximum terminal rotation time $t_{\theta_{max}}$ as 30s. We consider lasers to be operated in K-band, the power limit for satellites' terminal $P_{max}$ is set as 4 W \cite{mynaric}, the maximal equivalent isotropically radiated power (i.e. $P_{max}G_mG_n$) between any pair of satellites $m$ and $n$ as 53 dBW \cite{space}, the minimum data rate $C_{min}$ as 10 Kbps \cite{DBLP:journals/twc/MayorgaSP21}, thermal noise $\tau$ as 318 K \cite{temperature}, channel bandwidth $B$ as 15 Mbps \cite{DBLP:conf/icc/PiRWZZL22}, and carrier frequency $f$ as 23.28 GHz \cite{k}. The preference weights $\alpha$ and $\beta$ in Eq. \eqref{problem} is set to be ($\alpha$=$\frac{2}{3}$, $\beta$=$\frac{1}{3}$) and ($\alpha$=$\frac{1}{3}$, $\beta$=$\frac{2}{3}$) to reflect the transmission-energy-biased and the ISL-switching-cost-biased cases, respectively.

 \setlength{\tabcolsep}{3.8pt} 

\begin{table}[t]
    \centering
     \caption{Performance comparison between $\textsc{SatFlow}$ and benchmarks. ``/" separates the energy-consumption-biased and ISL-switching-cost-biased case. Note that SP-D’s low costs are at the expense of a much higher FVR.}
    \begin{tabular}{c|c|cccc}
    \hline
     Shell & Algorithm & Energy & ISL Switch. & Tot. Cost & FVR(\%)\\
     \hline
       \multirow{2}*{\makecell[c]{Starlink\\Shell A}} & {\makecell[c]{SP-F\&+Grid\\SP-D\&+Grid\\SP-F\&GEO\\SP-D\&GEO \\SP-F\&CapOpt\\SP-D\&CapOpt \\ SP-F\&RS\\SP-D\&RS \\ \textbf{\textsc{SatFlow}} }}  & {\makecell[c]{5.72\\0.37\\5.96\\0.37\\5.65\\0.34\\6.19/5.91\\0.35/0.40\\0.75/0.77}}  & {\makecell[c]{0\\0\\0.13\\0.13\\0.85\\0.85\\0.35/0.27\\0.35/0.27\\0.17/0.12}} & {\makecell[c]{3.82/1.91\\0.25/0.12\\4.02/2.08\\0.29/0.21\\4.05/2.45\\0.51/0.68\\4.25/2.15\\0.35/0.32\\0.56/0.34}} & {\makecell[c]{14.2\\14.2\\14.3\\14.3\\20.7\\20.7\\15.8/13.4\\15.8/13.4\\\textbf{4.1}/\textbf{3.8}}} \\
       \hline
       
        \multirow{2}*{\makecell[c]{Starlink\\Shell B}} & {\makecell[c]{SP-F\&+Grid\\SP-D\&+Grid\\SP-F\&GEO\\SP-D\&GEO \\SP-F\&CapOpt\\SP-D\&CapOpt \\ SP-F\&RS\\SP-D\&RS \\ \textbf{\textsc{SatFlow}} }}  & {\makecell[c]{5.24\\0.29\\5.34\\0.31\\5.74\\0.29\\5.61/5.22\\0.33/0.35\\0.57/0.58}}  & {\makecell[c]{0\\0\\0.21\\0.21\\0.71\\0.71\\0.32/0.28\\0.32/0.28\\0.16/0.11}} & {\makecell[c]{3.50/1.75\\0.20/0.10\\3.63/1.92\\0.28/0.24\\4.06/2.38\\0.43/0.57\\3.85//1.93\\0.32/0.30\\0.43/0.26}} & {\makecell[c]{15.4\\15.4\\16.2\\16.2\\26.0\\26.0\\18.7/16.4\\18.7/16.4\\\textbf{5.5}/\textbf{5.0}}} \\
       \hline
       
       \multirow{2}*{Kuiper} & {\makecell[c]{SP-F\&+Grid\\SP-D\&+Grid\\SP-F\&GEO\\SP-D\&GEO \\SP-F\&CapOpt\\SP-D\&CapOpt \\ SP-F\&RS\\SP-D\&RS \\ \textbf{\textsc{SatFlow}} }}  & {\makecell[c]{3.67\\0.22\\4.03\\0.58\\3.93\\0.22\\4.12/3.91\\0.66/0.87\\0.86/0.75}}  & {\makecell[c]{0\\0\\0.23\\0.23\\0.30\\0.30\\0.23/0.19\\0.23/0.19\\0.19/0.08}} & {\makecell[c]{2.45/1.22\\0.15/0.07\\2.76/1.50\\0.47/0.35\\2.72/1.51\\0.25/0.28\\2.82/1.43\\0.52/0.42\\0.64/0.30}} & {\makecell[c]{12.3\\12.3\\19.8\\19.8\\12.1\\12.1\\21.3/13.2\\21.3/13.2\\\textbf{6.3}/\textbf{5.4}}} \\
       \hline

    \end{tabular}

    \label{tab:overall}
\end{table}

\textbf{Traffic flow simulation.} 
To simulate dynamic flows, we select flows for each $\mathcal{T}_i$ according to a Youtube’s mobile streaming dataset \cite{youtube}.
It contains flow meta-information including source and destination addresses.
Google servers' locations are derived from their IP addresses \cite{ip}, while mobile users' locations are randomly assigned to cities with Google data centers \cite{google_center} due to users' anonymity. 
We then define three levels (as shown in Table \ref{tab:flow}) of traffic intensity according to each shell's estimated capacity for the usage of later simulation, reflecting various traffic heaviness. Flows' data rates are randomly assigned from 4, 6, and 8Mbp, which matches the recommendation for streaming videos of 720p, 960p, and 1080p, respectively \cite{bitrate}, reflecting the diversity of video types and customized visual quality.

\textbf{Hyper-parameters setting.} For \textit{SatFlow-L}, we set the training episode limit $k_{max}$ as 300 and the sub-gradient descent iteration limit $k_{sg}$ as 20. The penalty coefficient $\rho$ for the flow conservation violation is set as  1\textit{e}-10 to prioritize flow conservation. The update step size $\sigma$ is 1.9. For \textit{SatFlow-U}, the per-orbit group number $N_G$ is 4 for Starlink and 2 for Kuiper. The node embedding dimensions $S$, $C_0$, $C_1$ are set as 40, 32, and 16, respectively, and the MLP's hidden unit size $M_0$ is set as 20. We further set the maximal training episode $K_{max}$ as 1000, the update period of target $Q$ function as 5, the discounted factor $\phi$ as 0.95, and the batch size $N_{batch}$ as 256. We adopt the Adam optimizer with a learning rate of 0.0025.

\begin{figure*}[t]

\begin{minipage}[b]{.33\linewidth}
  \centering
\includegraphics[width=2.5in,height=1.8in]{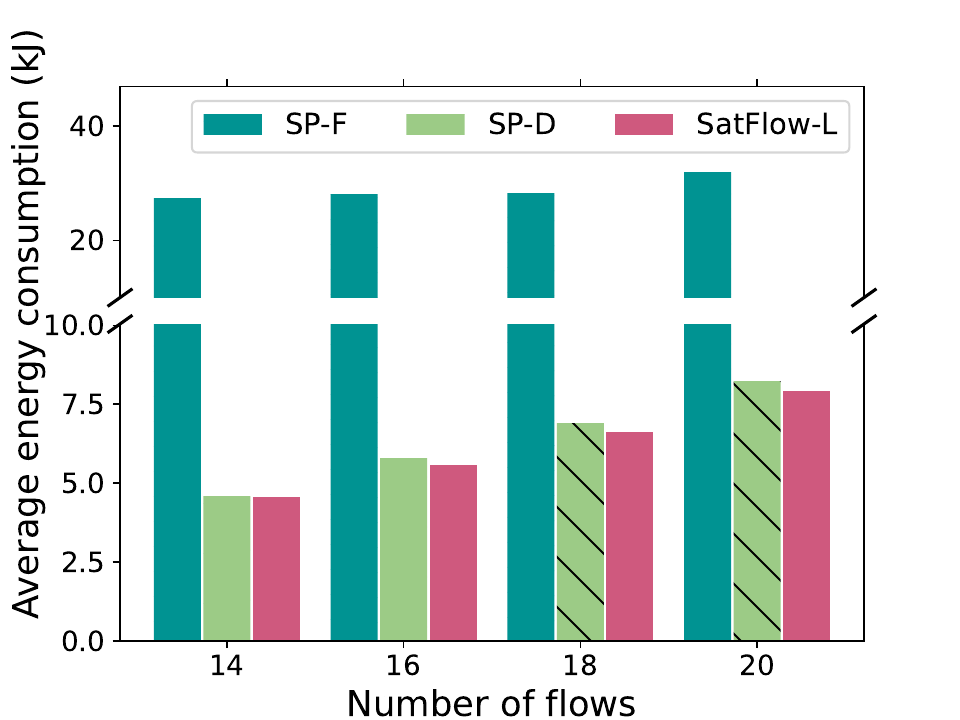}

  \centerline{(a) Starlink Shell A}\medskip
\end{minipage}
\begin{minipage}[b]{.33\linewidth}
  \centering
\includegraphics[width=2.5in,height=1.8in]{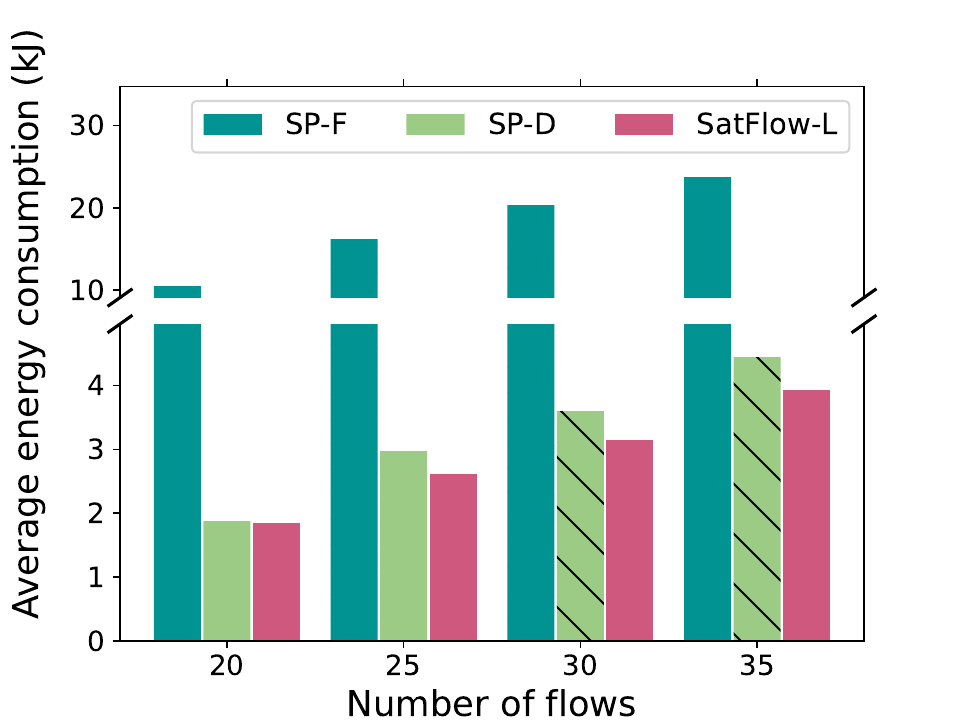}

  \centerline{(b) Starlink Shell B}\medskip
\end{minipage}
\hfill
\begin{minipage}[b]{.33\linewidth}
  \centering
\includegraphics[width=2.5in,height=1.8in]{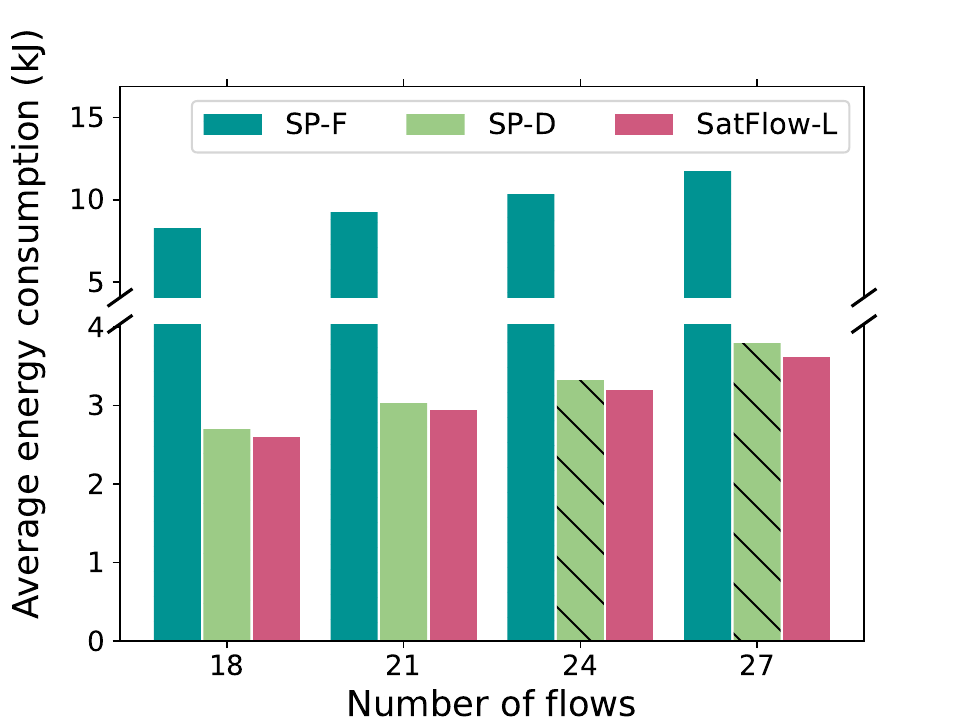}

  \centerline{(c) Kuiper}\medskip
\end{minipage}
\caption{Energy consumption comparison of various traffic and power allocation schemes.
}
\label{fig:short_a}
\end{figure*}

\begin{figure}[t]

  \centering
\includegraphics[width=2.5in,height=1.8in]{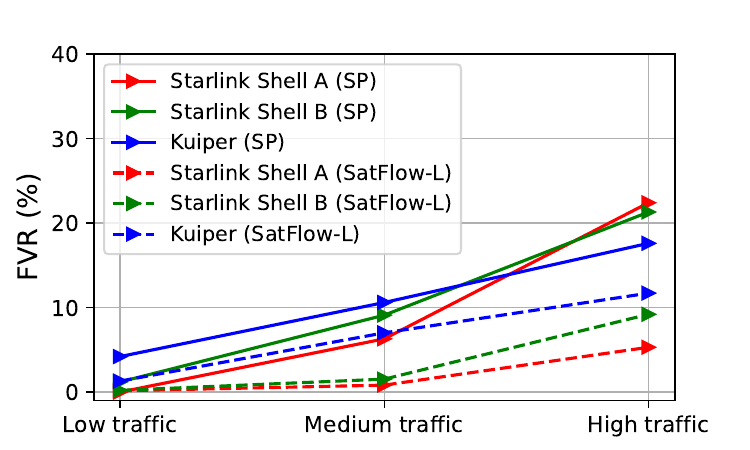}
\caption{FVR comparison of various traffic and power allocation schemes.
}
\label{fig:low_satsify}
\end{figure}

\begin{figure*}[th]

\begin{minipage}[b]{.33\linewidth}
  \centering
\includegraphics[width=2.5in,height=1.5in]{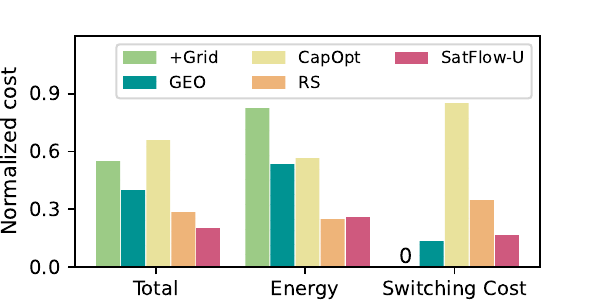}

  \centerline{(a) Starlink Shell A}\medskip
\end{minipage}
\begin{minipage}[b]{.33\linewidth}
  \centering
\includegraphics[width=2.5in,height=1.5in]{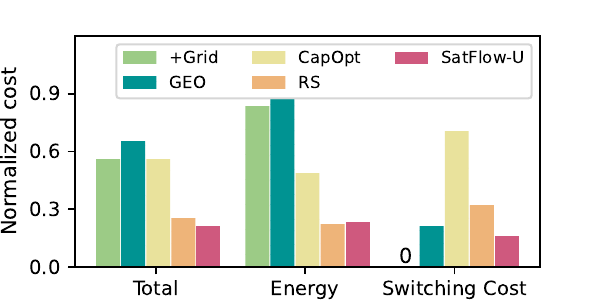}

  \centerline{(b) Starlink Shell B}\medskip
\end{minipage}
\hfill
\begin{minipage}[b]{.33\linewidth}
  \centering
\includegraphics[width=2.5in,height=1.5in]{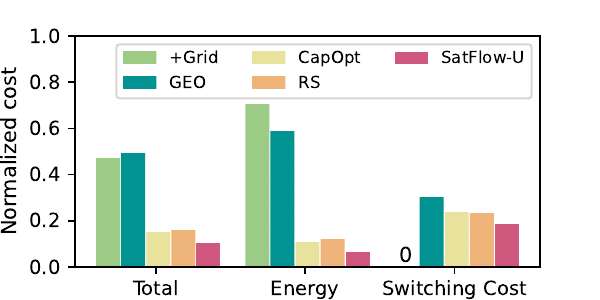}

  \centerline{(c) Kuiper}\medskip
\end{minipage}
\caption{Normalized cost comparison of various ISL re-establishment schemes for the terminal-energy-consumption-biased case. }
\label{fig:overall_e_biased}
\end{figure*}

\begin{figure*}[th]

\begin{minipage}[b]{.33\linewidth}
  \centering
\includegraphics[width=2.5in,height=1.5in]{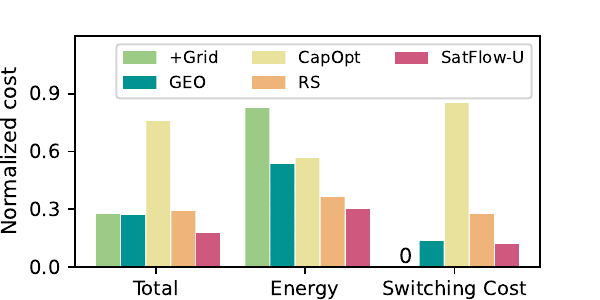}

  \centerline{(a) Starlink Shell A}\medskip
\end{minipage}
\begin{minipage}[b]{.33\linewidth}
  \centering
\includegraphics[width=2.5in,height=1.5in]{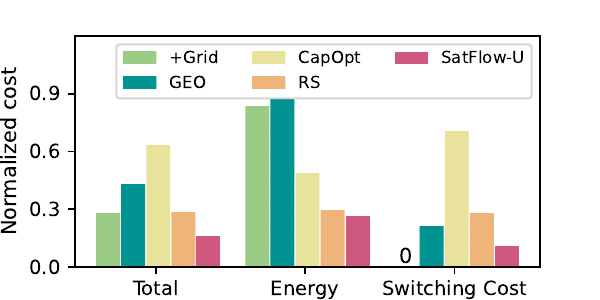}

  \centerline{(b) Starlink Shell B}\medskip
\end{minipage}
\hfill
\begin{minipage}[b]{.33\linewidth}
  \centering
\includegraphics[width=2.5in,height=1.5in]{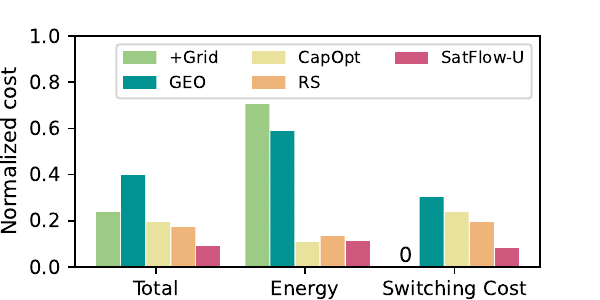}

  \centerline{(c) Kuiper}\medskip
\end{minipage}
\caption{Normalized cost comparison of various ISL re-establishment schemes for the ISL-switching-cost-biased case.
}
\label{fig:overall_sw_biased}
\end{figure*}

\subsection{Benchmarks}

 We identify the following state-of-the-art algorithms to allocate terminal power for ISL establishment and traffic: 

    \textit{1) SP-F}: The allocated power for each ISL is either zero or $P_{max}$ between adjacent $t\in\mathcal{T}_i$, depending on whether the ISL is within the output of a flow's path obtained from shortest path algorithm, which is commonly used for routing in LEO networks \cite{DBLP:conf/infocom/LaiLWLXW22}\cite{DBLP:conf/hotnets/Handley18}. For better energy saving, we use the extra terminal energy cost for routing a flow as the edge weight.
    
    \textit{2) SP-D}: It differs from \textit{SP-F} by allowing a dynamic and continuous-valued power allocation scheme, which takes the minimum power required for the allocated data rate.

We identify the following state-of-the-art algorithms to conduct ISL re-establishment.

  \textit{1) +Grid} \cite{grid}: A static matching scheme that satellites establish inter-plane ISLs with the ones of zero index offset.

 \textit{2) Geographical matching algorithm (GEO)} \cite{917071}: A classic matching scheme that latitude is evenly divided into $N_{spo}$ areas, and satellites establish ISLs with those in the same area. 
 
 \textit{3) Capacity-optimal algorithm (CapOpt)} \cite{DBLP:journals/twc/MayorgaSP21}: It always chooses the plan leading to the largest average link capacity, aiming to increase ISLs' cost efficiency for transmission.

  \textit{4) Random search (RS)}: a searching-based method that outputs the smallest total costs among several randomly-generated ISL re-establishment schemes. For a fair comparison, the searching iteration is set as the same as ours.

Note that we did not take traditional single-agent-RL based method (such as \cite{DBLP:conf/sigcomm/ZhuGATZJ21}) as one of our benchmarks since we find it does not converge after a large number of steps.

\subsection{Performance Metrics}
We consider the following metrics for evaluation.

 \textit{1) Average energy consumption}: It refers to the average per-satellite energy consumption during a period.
 
  \textit{2) Average flow violation ratio (FVR)}: It reflects the conformity of the flow conservation constraint (i.e. Constraint \eqref{c2}) during a period. For period $\mathcal{T}$, we define FVR as follows to ensure that it equals 1 when no flows are routed:

  \begin{equation}
      \frac{1}{2\lvert\Omega_t\rvert\lvert \mathcal{T}\rvert}\sum_t^{\mathcal{T}}\sum_{\omega\in\Omega_t}\sum_{m\in\mathcal{M}}\sum_{n\in\mathcal{N}_m}\left\lvert\frac{Y_{m,n}^{\omega,t}-Y_{n,m}^{\omega,t}-[\textbf{b}^\omega]_m}{d_{\omega}}\right\rvert
  \end{equation}

  \textit{3) Normalized cost}: It reflects the normalized cost including the terminal energy consumption cost, the ISL switching cost, and the total cost. 
 The total cost is the weighted average of the former two costs, parameterized by $\alpha$ and $\beta$ in Eq. \eqref{problem}.

\subsection{Experiment Results}

\subsubsection{Overall performance}

We compare \textsc{SatFlow} with various benchmarks, shown in Table \ref{tab:overall}. Since we observe that the traditional fixed power allocation scheme consumes much more energy than the dynamic one, we normalize the energy and ISL switching costs to be within [0, 1] under the dynamic power allocation schemes to better reflect the differences between various schemes. To reflect the generality over various traffic intensities, we randomly select a flow number $k$ from each traffic intensity level to generate three traffic scenarios, run the algorithms on them, and record the average performance.  We find that \textsc{SatFlow} reduces FVR by up to 21.0\%. Furthermore, the normalized total costs are significantly optimized compared with those with the traditional fixed allocation scheme SP-F by up to 89.4\%. Even though SP-D could consume smaller normalized costs than \textsc{SatFlow} by up to 16.2\%, it sacrifices for much flow drop, thus leading to a much larger FVR by up to 21.0\%. We further compare the flow throughput across the constellations, and find that \textsc{SatFlow}'s throughput is 177.2, 339.3, and 227.8 Mbps for each shell respectively, increasing by 8.2\%, 10.1\%, and 9.4\% compared with the other various benchmarks on average.

\subsubsection{Lower-level Traffic and Power Allocation Performance}

We then perform an ablation study to uncover the effects of both our lower-level and upper-level modules.
We first analyze the effect of our lower-level module on energy savings. Without loss of generality, we run the module on a randomly generated topology during the first period of length $D_i$. We compare various traffic and power allocation methods regarding the average energy consumption, shown in Fig. \ref{fig:short_a}. For fair comparisons, we choose the number of flows $k$ within a range that no flows are dropped under SP-F and SP-D, and we run our module until FVR is less than 1\textperthousand. We find that the traditional fixed power allocation scheme SP-F consumes 3.2 times more energy than SP-D on average, and \textit{SatFlow-L} consumes up to 21.3$\%$ less energy than SP-D further since it considers detailed concurrent flows in the network. Furthermore, we observe that the margin gets wider when $k$ is larger. 
We then set $k_{max}$ back to 300, and compare \textit{SatFlow-L} with benchmarks regarding FVR under various traffic intensities, as shown in Fig. \ref{fig:low_satsify}. We find that \textit{SatFlow-L} induces $1.8\%\sim 17.1\%$ lower FVR, and the superiority increases when the traffic intensity increases.

\subsubsection{Upper-level ISL Re-establishment Performance}

To study the advantage of \textit{SatFlow-U} on cost savings, we restrict all the benchmarks to use our \textit{SatFlow-L} algorithm for the lower-level module, and compare \textit{SatFlow-U} with various ISL re-establishment benchmarks, as shown in Fig. \ref{fig:overall_e_biased} and Fig. \ref{fig:overall_sw_biased}, with the former figure depicts the transmission-energy-biased case, while the latter one captures the ISL-switching-cost-biased case. To better exhibit the difference between various schemes, we normalize energy and ISL switching costs within [0, 1] under the usage of \textit{SatFlow-L}. We randomly select $k$ from each traffic intensity level for three times and record the average performance. We find that \textit{SatFlow-U} consumes the least total cost under both biased cases and all the shells, with an average of $23.6\%$, $28.2\%$, $33.1\%$, and $8.2\%$ reduction compared with +Grid, GEO, CapOpt, and RS, respectively.

To be specific, for the energy consumption, both +Grid and GEO aim at reducing the inter-satellite latitude difference, which nevertheless could not guarantee high link cost efficiency due to satellites' spherical orbits, thus consuming $59.0\%$ and $46.6\%$ more energy on average than our method, respectively. Although CapOpt considers link cost efficiency and thus narrows the gap of energy consumption with ours to $18.4\%$ on average, it neglects the end-to-end performance. For the normalized ISL switching cost, we find that +Grid consumes no such cost due to its static topology. GEO consumes a relatively stable normalized switching cost, swinging between 0.13 to 0.30, while CapOpt could incur a cost of up to 0.85 due to the constellation dynamics. Finally, we find that \textit{SatFlow-U} explores solutions more efficiently than RS, with a normalized total cost reduction of up to 11.8\%.

\section{Conclusion}
\label{sec:conc}
The recent surging construction of LEO mega-constellations has made global communication with low latency possible.  Nevertheless, it poses significant challenges for network planning on these mega-constellations due to their scale and dynamics. To cope with these challenges, we propose \textsc{SatFlow}, a hierarchical distributed framework to determine network topology, traffic allocation, as well as terminal power. 
 \textsc{SatFlow}'s upper-level module utilizes a multi-agent RL to make long-term ISL re-establishment plans, while its lower-level module uses distributed Lagrangian optimization to determine traffic and terminal power allocation. Extensive simulations on various mega-constellations validate the great scalability of \textsc{SatFlow} on the constellation size, with up to 21.0\% drop for flow violation ratio and up to an 89.4\% reduction on total network operational costs.

\newpage
\appendices

\bibliography{ref.bib}
\bibliographystyle{IEEEtran}

\end{document}